\documentclass{egpubl}
 
\JournalPaper         
\usepackage[T1]{fontenc}
\usepackage{dfadobe}  

\biberVersion
\BibtexOrBiblatex
\usepackage[backend=biber,bibstyle=EG,citestyle=alphabetic,backref=true,giveninits=true]{biblatex} 

\AtEveryBibitem{%
\ifentrytype{book}{
    \clearfield{url}%
    \clearfield{urlyear}%
    \clearfield{issn}%
    \clearfield{review}%
    \clearfield{series}
}{}
\ifentrytype{collection}{
    \clearfield{url}%
    \clearfield{urlyear}%
    \clearfield{issn}%
    \clearfield{review}%
    \clearfield{series}
}{}
\ifentrytype{incollection}{
    \clearfield{url}%
    \clearfield{urlyear}%
    \clearfield{issn}%
    \clearfield{review}%
    \clearfield{series}
}{}
\ifentrytype{article}{
    \clearfield{url}%
    \clearfield{urlyear}%
    \clearfield{issn}%
    \clearfield{review}%
    \clearfield{series}
}{}
\ifentrytype{inproceedings}{
    \clearfield{url}%
    \clearfield{urlyear}%
    \clearfield{issn}%
    \clearfield{review}%
    \clearfield{series}
}{}
}

\electronicVersion
\PrintedOrElectronic
\ifpdf \usepackage[pdftex]{graphicx} \pdfcompresslevel=9
\else \usepackage[dvips]{graphicx} \fi

\usepackage{egweblnk} 

\title[Visual Parameter Selection for Spatial BSS]%
      {Visual Parameter Selection for Spatial Blind Source Separation}

\author[N.~Piccolotto et al.]
{\parbox{\textwidth}{\centering
N. Piccolotto$^{1}$\orcid{0000-0001-6876-6502},
M. Bögl$^{1}$\orcid{0000-0002-8337-4774},
C. Muehlmann$^{2}$\orcid{0000-0001-7330-8434},
K. Nordhausen$^{3}$\orcid{0000-0002-3758-8501},
P. Filzmoser$^{2}$\orcid{0000-0002-8014-4682},
 and S. Miksch$^{1}$\orcid{0000-0003-4427-5703}
        }
        \\
{\parbox{\textwidth}{\centering $^1$TU Wien, Institute of Visual Computing and Human-Centered Technology, Austria\\
         $^2$TU Wien, Institute of Statistics and Mathematical Methods in Economics, Austria\\
        $^3$ University of Jyväskylä, Finland
       }
}
}

\usepackage{csquotes}
\usepackage[plain]{fancyref}
\usepackage{xcolor}
\usepackage{bm}
\usepackage{MnSymbol}
\usepackage{amsfonts}
\usepackage{amsmath}
\usepackage{booktabs}
\usepackage{caption}
\usepackage{subcaption}

\usepackage{pgf}					
\usepackage{tikz}					
\usepackage{varwidth}
\usetikzlibrary{calc}
\usetikzlibrary{%
  arrows,%
  shapes,
  chains,%
  matrix,%
  positioning,
  scopes,%
  decorations.pathmorphing,
  shadows,%
  calc
}\tikzset{
     textWcolor/.style={
        		circle,
            fill=white, 
            draw=black, 
            line width=1pt,
            font=\sffamily\scriptsize,
            opacity=0.7,
            inner sep=0pt,
            minimum size=10pt
    },
    textWborder/.style={
        		circle,
            fill=none,
            draw=black, 
            line width=1pt,
            font=\normalfont\bfseries\small,
            opacity=0.7,
            inner sep=0pt,
            minimum size=12pt
    },
    textOnly/.style={
        		rectangle,
        		fill=none, 
            font=\normalfont\scriptsize,
            inner sep=0pt,
            minimum size=10pt
    },
    cycleLabels/.style={
        		rectangle,
            fill=none,
            draw=none,
            font=\normalfont\small,
            inner sep=0pt,
            anchor=west
    },
    mathnode/.style={
        		rectangle,
            fill=none,
            draw=none,
            font=\normalfont\normalsize,
            inner sep=0pt,
            anchor=west
    },
    sfnode/.style={
        		rectangle,
            fill=none,
            draw=none,
            font=\sffamily\small,
            inner sep=0pt,
            anchor=west
    },
    image overlay index/.style={
        		circle,
            fill=col2,
            draw=black!40,
            line width=0.5pt,
            font=\normalfont\scriptsize,
            opacity=0.7,
            inner sep=0pt,
            minimum size=10pt
    }
}

\begin{document}


\maketitle
\begin{abstract}
   Analysis of spatial multivariate data, i.e., measurements at irregularly-spaced locations, is a challenging topic in visualization and statistics alike. Such data are integral to many domains, e.g., indicators of valuable minerals are measured for mine prospecting. Popular analysis methods, like PCA, often by design do not account for the spatial nature of the data. Thus they, together with their spatial variants, must be employed very carefully. Clearly, it is preferable to use methods that were specifically designed for such data, like spatial blind source separation (SBSS). However, SBSS requires two tuning parameters, which are themselves complex spatial objects. Setting these parameters involves navigating two large and interdependent parameter spaces, while also taking into account prior knowledge of the physical reality represented by the data. To support analysts in this process, we developed a visual analytics prototype. We evaluated it with experts in visualization, SBSS, and geochemistry. Our evaluations show that our interactive prototype allows to define complex and realistic parameter settings efficiently, which was so far impractical. Settings identified by a non-expert led to remarkable and surprising insights for a domain expert. Therefore, this paper presents important first steps to enable the use of a promising analysis method for spatial multivariate data.


\end{abstract}
\section{Introduction}
\label{sec:introduction}

Many domains work with multivariate quantitative measurements at different locations, i.e., multivariate spatial data. Such data can stem from, e.g., geochemical analyses of soil samples for the purpose of mine prospecting \cite{haldar2018a} or investigations of environmental pollution \cite{reimann2014}. Depending on the specific goal and application, various tasks need to be carried out on such a spatial dataset, like dimension reduction (DR), or finding meaningful linear combinations of involved variables \cite{BaileyKrzanowski2012, Wackernagel2003}. Spatial blind source separation (SBSS) \cite{nordhausen2015, bachoc2020, muehlmann2021} is specifically designed for multivariate spatial data and reveals linear combinations of such data.  It brings various advantages compared to alternative methods (\Fref{sec:related-work--spatial-statistics}), e.g., it keeps the well known loadings-scores scheme from principal component analysis and properly accounts for spatial dependence due to its model-based approach. Therefore, latent dimensions identified with SBSS often correspond to the physical reality where data was collected, making it a superior analysis tool for spatial data. When irrelevant dimensions are discarded, SBSS serves as DR method as well. SBSS has been successfully applied to a geochemical dataset \cite{nordhausen2015} and may be potentially used in any application domain that involves multivariate quantitative measurements at different locations.

\begin{figure}[h]
     \centering
     \begin{subfigure}[b]{0.115\textwidth}
         \centering
         \includegraphics[width=\textwidth]{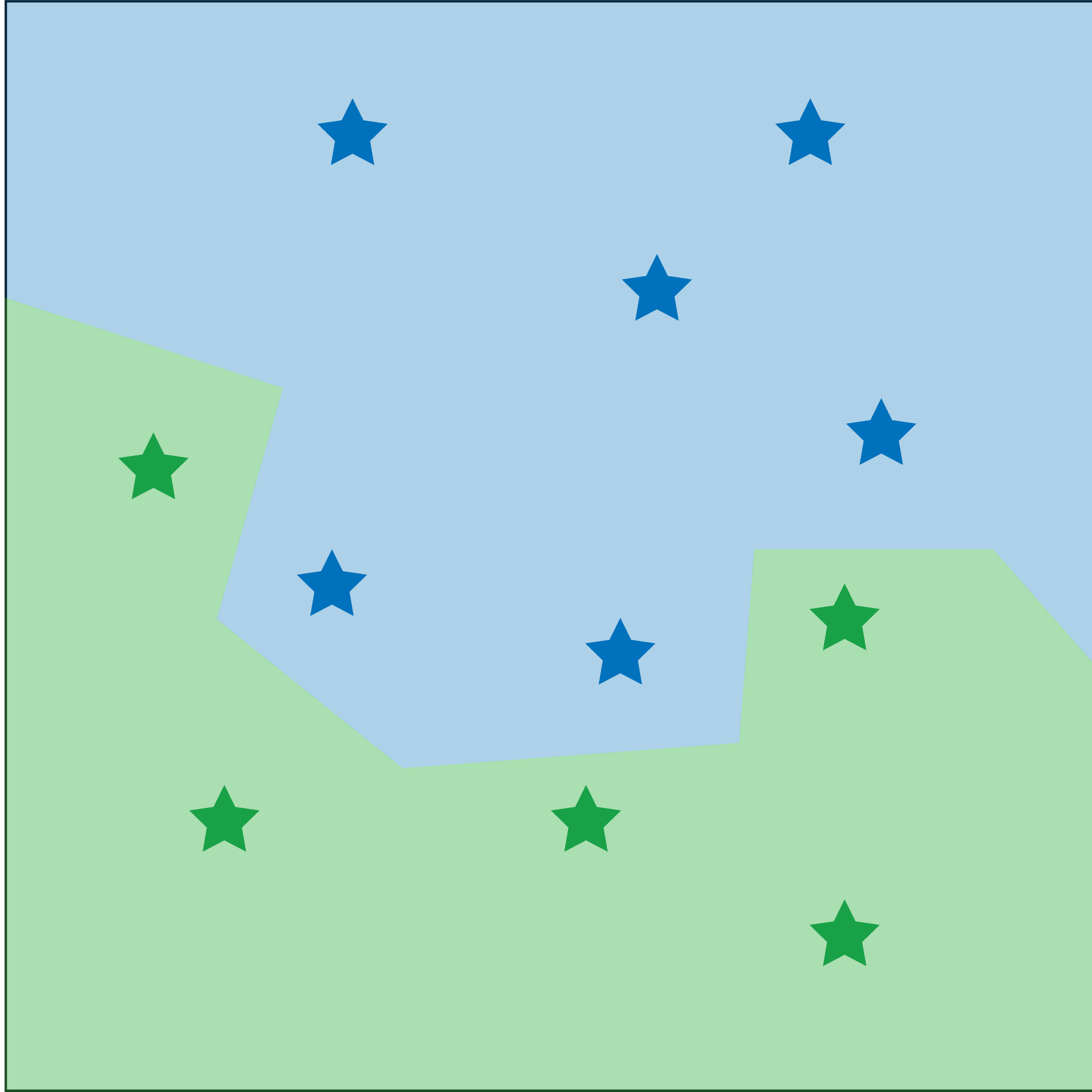}
         \caption{}
         \label{fig:parameters-regionalization}
     \end{subfigure}
     \begin{subfigure}[b]{0.115\textwidth}
         \centering
         \includegraphics[width=\textwidth]{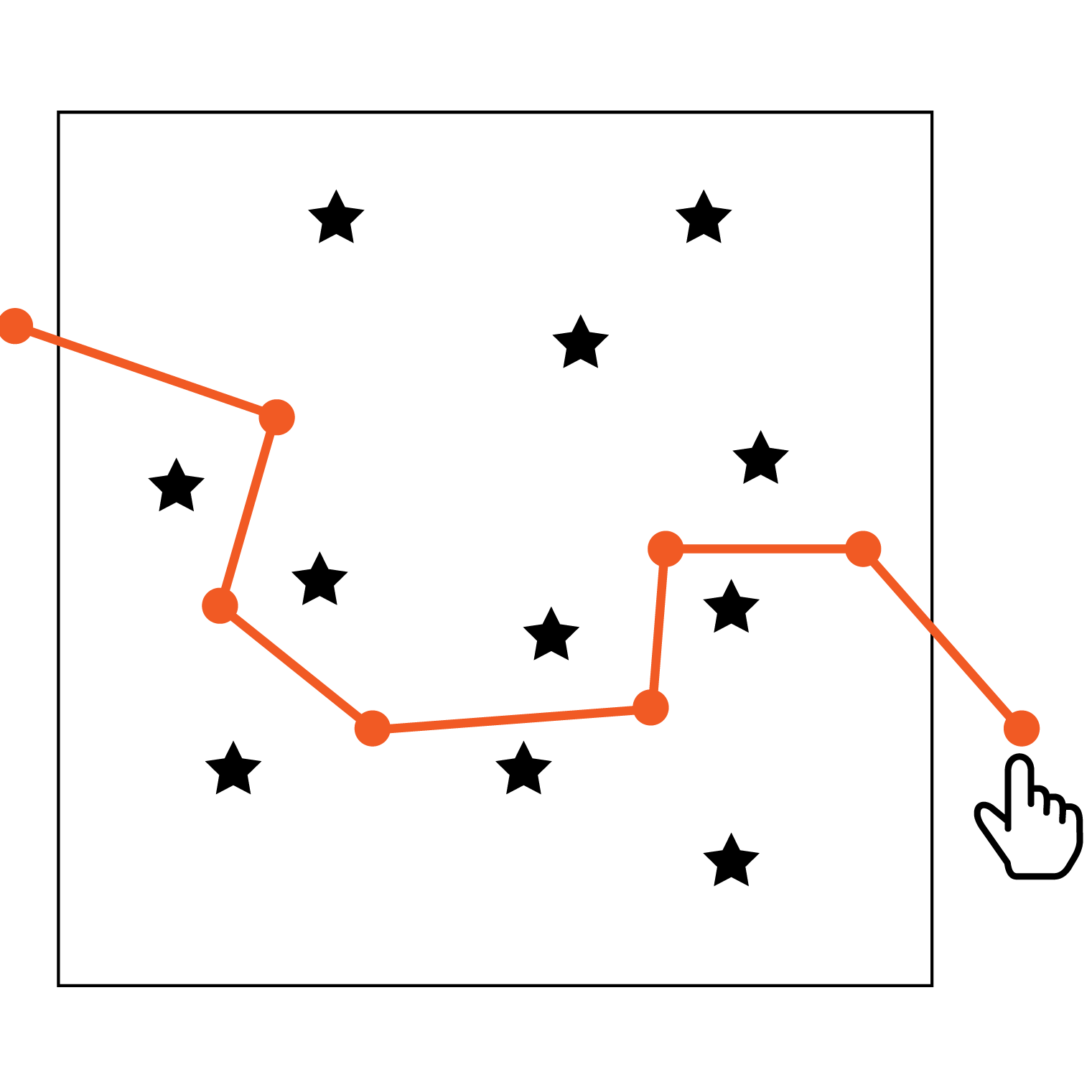}
         \caption{}
         \label{fig:parameters-ixn-region}
     \end{subfigure}
     \begin{subfigure}[b]{0.115\textwidth}
         \centering
         \includegraphics[width=\textwidth]{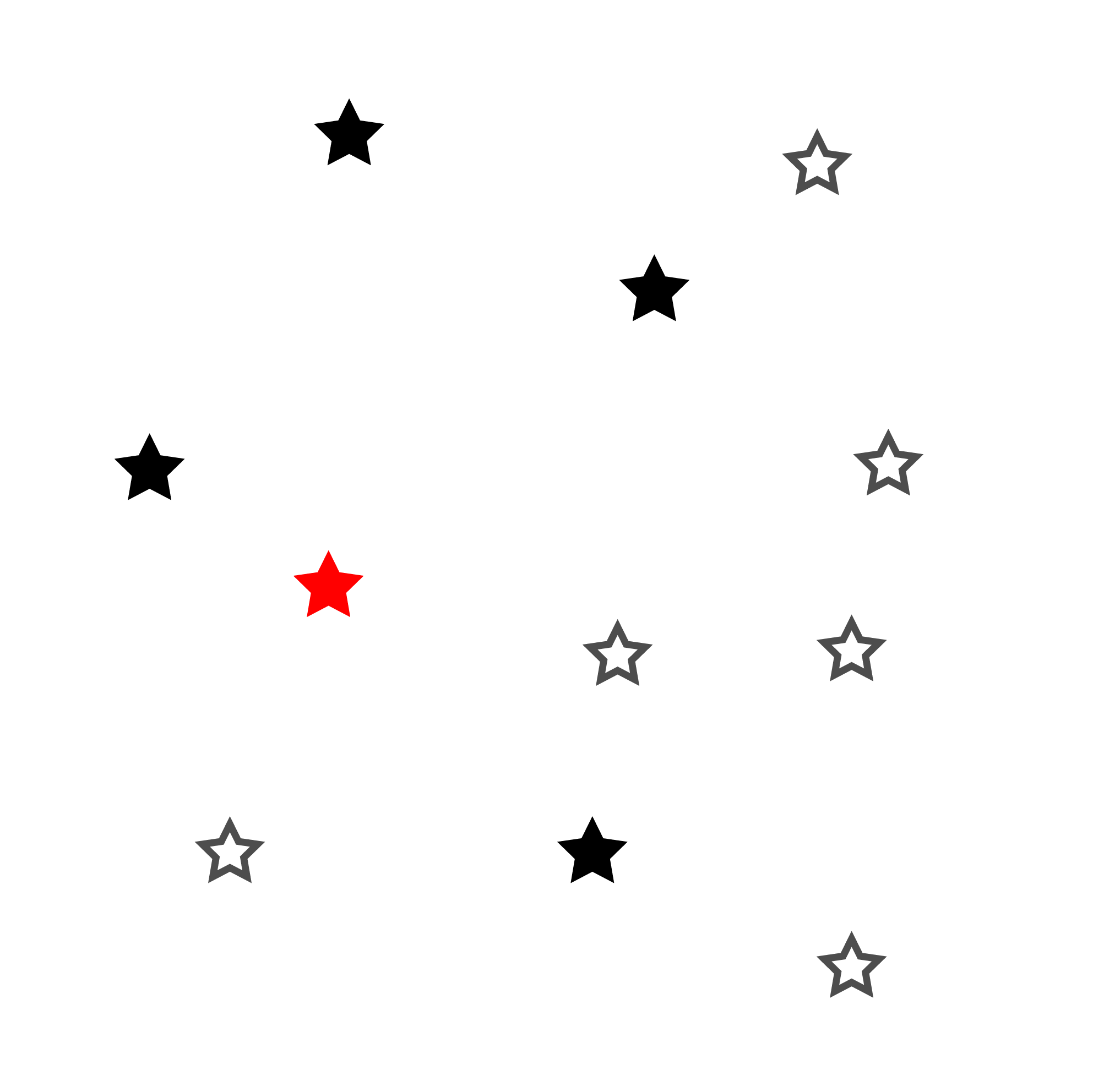}
         \caption{}
         \label{fig:parameters-kernel}
     \end{subfigure}
     \begin{subfigure}[b]{0.115\textwidth}
         \centering
         \includegraphics[width=\textwidth]{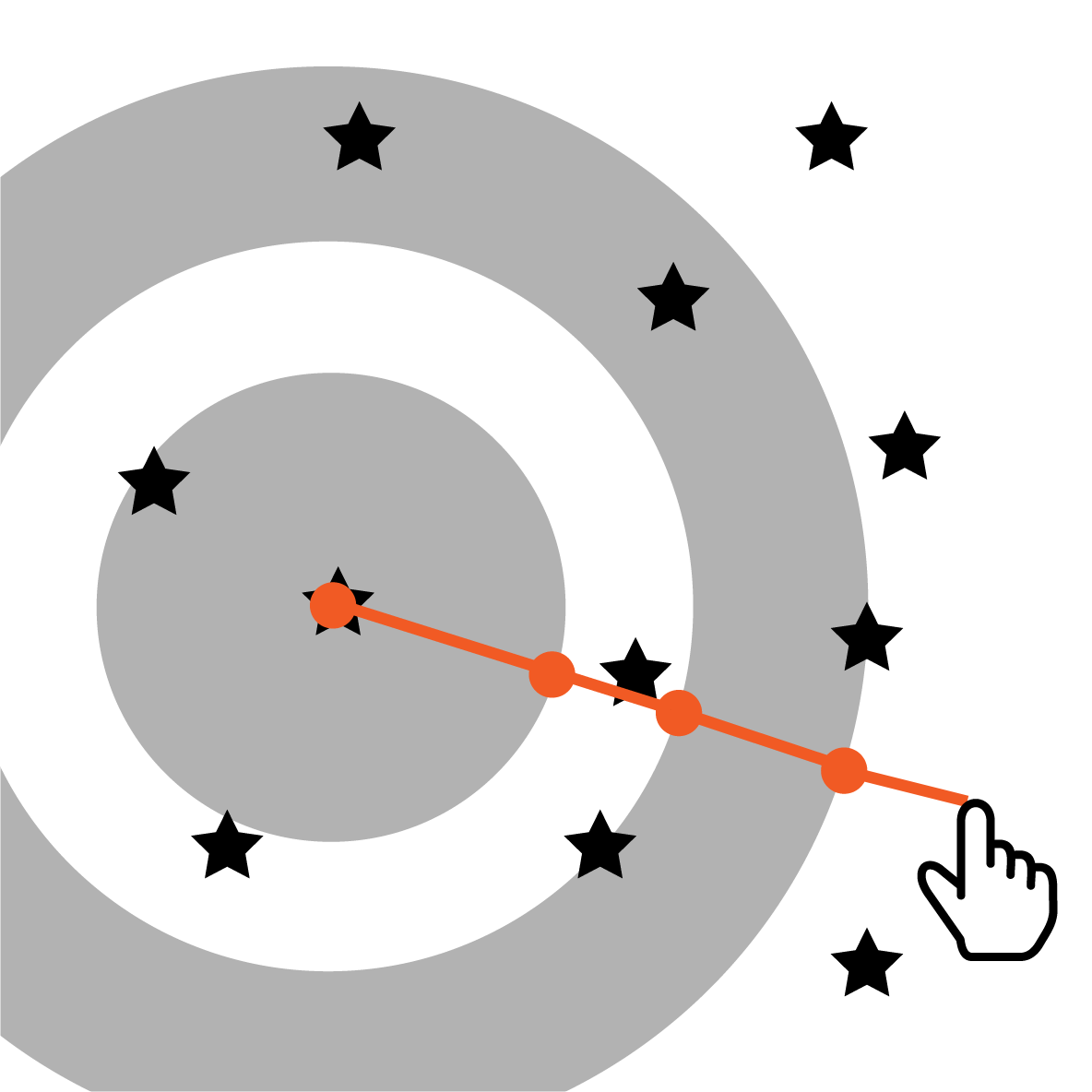}
         \caption{}
         \label{fig:parameters-ixn-kernel}
     \end{subfigure}
    \caption{SBSS parameters illustrated on the same locations. A regionalization (a) into a green and blue region. Two ring-shaped kernels (c) as applied to the red location. Black locations are the red location's neighbourhood. Our prototype allows setting those parameters with direct manipulation by splitting a region along a polyline (b) and choosing kernel radii (d).}
    \label{fig:parameters}
\end{figure}

However, a challenge to the effective use of SBSS in practice are two spatial tuning parameters that need to be set: A partition of the spatial domain into non-overlapping regions, and a configuration of non-overlapping ring-shaped kernels (\Fref{fig:parameters}, see \Fref{sec:related-work--spatial-statistics}). The performance of SBSS depends largely on the choice of these tuning parameters, but the size of the parameter space is overwhelming for analysts. Theoretical guidelines about the optimal choice of tuning parameters exist (see \Fref{sec:parameter-considerations}), but they leave plenty of room for human judgement and automatic optimization does not seem feasible. Further complicating the issue, the current tool of analysts is text-based and not well suited to support them in their tasks.

As with many data analysis methods, the expertise of the human analyst is vital to SBSS parameter selection. We believe that visual analytics (VA) \cite{thomas2005, keim2008} can enable the effective use of SBSS in practice. VA pairs automatic data analysis with visual methods to combine and thereby enhance the computer's and human's individual strengths. To this end, we designed and developed interactive visualizations in collaboration with SBSS experts, who are co-authors of this paper, and an expert in geochemistry. We evaluated our approach with five visualization experts using a heuristic for value-driven visualization \cite{wall2019} and one domain as well as two external SBSS experts.

Our contributions are the following:

\begin{itemize}
    \item A task description for SBSS parameter selection,
    \item a visualization design to support parameter selection for SBSS, including novel and existing interactions and visualizations,
    \item an evaluation of the design with experts in visualization, SBSS, and geochemistry.
\end{itemize}

Our contributions are interesting and relevant for the visualization community. They represent an important first step to enable the use of SBSS, a desirable multivariate spatial analysis method. Blind source separation and geostatistics in general and SBSS especially have been little explored so far. In the context of visual parameter analysis our scenario is notable because of spatial tuning parameters.



\section{Related Work}
\label{sec:related-work}

In the following we describe spatial statistics (\Fref{sec:related-work--spatial-statistics}), visualizations for geospatial point data (\Fref{sec:related-work--geospatial-vis}) and interactions with parameter spaces (\Fref{sec:related-work--parameter}) to contextualize our approach.

\subsection{Spatial Statistics}
\label{sec:related-work--spatial-statistics}

Geostatistics is concerned with the analysis of data that show a natural order in space. Typically, many measurements at different sample locations are taken and the main source of information for proper statistical analysis of such multivariate spatial data is given by spatial dependence (cf. Tobler's law). Geostatisticians are faced with a wide variety of tasks, e.g., predicting the data at unobserved sample locations, dimension reduction, or finding meaningful linear combinations \cite{BaileyKrzanowski2012, Wackernagel2003}. Proper modelling of the spatial dependence is crucial for them. In the geostatistical framework it is assumed that the spatial data at hand are generated by a family of $p$-variate random vectors $\bm x (\bm s) = (x_1 (\bm s), \dots, x_p (\bm s))^\top$ indexed by elements $\bm s$ of the so-called spatial domain $S \subseteq \mathbb{R}^2$, e.g., longitude-latitude coordinates. Such a family of random vectors is referred to as multivariate random field. Spatial dependence is characterized by the so-called spatial covariance matrix which evaluates the covariance between the random field at two different sample locations. Often, the semi-variogram (covariance of the difference processes) is used in favor of the spatial covariance as it avoids the estimation of the mean but is usually harder to interpret. Modeling of proper covariance matrix functionals is a demanding task and usually simplified by further assumptions \cite{GentonKleiber2015}. The second-order stationary assumption yields that the spatial covariance is invariant under shifts, i.e., the spatial covariance is the same for the whole field and only dependent on the distance between two sample locations. In contrast, the spatial covariance function of a non-stationary random field depends on specific locations and distances between locations and is therefore usually much more demanding to model.

We will outline the advantages of SBSS over principal component analysis (PCA) and its spatial variants \cite{Jolliffe1986,Demsar2013}, because they are well known and widely used. For an overview of geostatistical methods see, e.g., \cite{BaileyKrzanowski2012, Wackernagel2003}. The classical PCA finds orthogonal directions of the data that maximize variance. It does so by the eigen-decomposition of the covariance matrix $\mathbf{Cov}$, yielding the orthogonal loadings matrix $\mathbf{U}$ and uncorrelated principal components (scores) $\mathbf{U} \mathbf{x}$. Two variants for spatial data are considered in the literature, both use the same methodology as classical PCA, but adapt $\mathbf{Cov}$. The so-called geographically weighted PCA \cite{fotheringham2002,HarrisEtAl2015} uses spatial information implicitly as it computes multiple $\mathbf{Cov}$ for each sample location based on the neighbors. This leads to local PCA solutions and different loadings for each sample location, which is very time-consuming to interpret. Another variant diagonalizes the product of $\mathbf{Cov}$ and a measure of spatial dependence (Moran's~I) \cite{JombartEtAl2008}, which leads to a trade-off between maximum spatial dependence and maximum marginal covariance in components. It is not clear which properties in terms of spatial/marginal dependence these components actually show. Generally, the advantage of PCA are feasible interpretations of the results in terms of the loadings-scores scheme. However, PCA and both its spatial variants lack a statistical model, therefore, is not clear which spatial and/or marginal dependence properties the results actually have. SBSS, on the other hand, provides both and can find physically meaningful processes which generated the data that also have certain well-defined statistical properties.


In recent literature \cite{nordhausen2015,bachoc2020,muehlmann2021} the methodology of blind source separation (BSS) \cite{ComonJutten2010} was combined with principles of stationary/non-stationary spatial data analysis, resulting in spatial blind source separation (SBSS) for stationary and non-stationary source separation (SNSS) for non-stationary spatial data. For simplicity, both versions are referred to as SBSS in this paper. The SBSS framework is appealing as it keeps the advantageous loadings-scores interpretation scheme but finds the solution by specifically accounting for spatial dependence, as it is mainly designed to find physically meaningful components. Moreover, SBSS does not restrict the loadings matrix to be orthogonal as PCA does. More meaningful components of a geochemical dataset were found in comparison to PCA by a domain expert \cite{nordhausen2015}, and pre-processing the data with SBSS in spatial prediction tasks simplifies the task but keeps the performance compared to other methods \cite{muehlmann2020}. The SBSS loadings matrix $\mathbf{W}$---often denoted as unmixing matrix---is found by jointly diagonalizing $\mathbf{Cov}$ and a number of so-called local covariance matrices $\mathbf{LCov}$ leading to the random field $\mathbf{W} \bm x (\bm s)$ (latent field) consisting of uncorrelated and spatially uncorrelated components. Local covariance matrices are computed by a weighted average of the spatial covariance matrix for all pairs of sample locations in a part of the spatial domain (regionalization, compare \Fref{fig:parameters-regionalization}). The weights are determined by a kernel which only accounts for sample location pairs that are at least separated by $r_{in}$ and at most separated by $r_{out}$ (compare \Fref{fig:parameters-kernel}). A regionalization is needed to account for non-stationarity of the random field, while the kernels specify local proximity and attempt to measure spatial dependency. Thus, for stationary data one region is sufficient and if there is no spatial dependency present the kernels are not informative.

The crucial point which determines the performance of the SBSS methods is the choice of $\mathbf{LCov}$ matrices or more precisely choosing a set of radii parameters (kernels) and a suitable domain subdivision (regionalization). Theoretical guidelines hint that theses parameters should be chosen such that the spatial dependence of the latent field components is as different as possible. However, the practical usefulness of this statement is limited as the latent field is unknown a-priori, which opens the door for parameter selection supported by sophisticated visual analytic methods.

\subsubsection{Spatial Data Analysis with Statistics and Visualization}

After the influential work by Cleveland and McGill on graphical perception and dynamic graphics in the 1980s, researchers started to apply these ideas to spatial data. Haslett et al. \cite{haslett1991} used coordinated multiple views with interactive highlighting to find anomalies in a geochemical dataset. The linked views in question included dynamic statistical graphics, such as a variogram cloud \cite{cressie1993m}, histograms, and a scatterplot matrix, as well as geographic views (a map). GeoVISTA Studio \cite{gahegan2002}, a visual programming environment for spatial data analysis, extended this approach and combined state-of-the-art visualizations with statistical methods for, e.g., classification. Demšar et al. \cite{demsar2008} used similar dynamic graphics but to explore spatially varying parameters of geographically weighted regression instead of the original spatial dataset. Dykes and Brunsdon \cite{dykes2007} suggested adjustments to well known statistical graphics to make them work in a geographically weighted setting and for multiple spatial scales. The latter was also a focus for Goodwin et al. \cite{goodwin2016}, who, more recently, used local regression coefficients to guide the analysis of a spatial dataset. To summarize, previous efforts have been put into using visualization to enable spatial exploration of the \emph{outcome} of statistical methods. While that is future work we plan, this paper aims to enable the use of a spatial analysis method in the first place.

\subsection{Visualization of Geospatial Point Data}
\label{sec:related-work--geospatial-vis}

As we see it necessary to visualize multivariate 2D spatial point data to facilitate SBSS parameter selection, the visualization of spatial point data is related to our work. Point data is quite common in geospatial visualizations.
When the interest is in a variable's value, dot maps are often used. In those, each point is represented by one visual mark, like a circle. Other visual variables are used to encode the actual value, such as area or color. Issues may occur, e.g., when the data distribution has long tails (common in geochemistry), as a few extreme values then reduce perceptual accuracy for the majority of data points. Zhou et al. proposed the point grid map \cite{zhou2017}, in which visual marks are aligned on a grid such that directional relations are preserved. Typographic properties, like font weight, as visual channels have been explored by Brath and Bassini \cite{brath2017}. When there is little space for individual marks, pixel maps \cite{keim2000} are an option. However, these approaches distort the location of points, which is crucial information in our case. Heatmaps and isocontours are employed when the number of points is too big for individual marks. On irregular points, like in our case these do, however, require some preprocessing as variable values need to be interpolated or resampled onto a regular grid.

There are also approaches to present point value without per-point marks on a map. Turkay et al. \cite{turkay2014} proposed attribute signatures, in which the analyst draws a path through space and connected small multiple line charts show the value of variables along the path. Their approach scales to many variables, but only shows a small portion of variable values. Bouts et al. \cite{bouts2016} warped the geographic space such that points with similar value are moved near each other, an idea from DR. While an interesting idea, we believe it would be unintuitive for our anticipated users.

Heatmaps are also useful to show the density of points, when individual marks tend to overlap. In this case, some abstraction is necessary. When the points are located along a road network, visual marks can be encoded along the streets with bristle maps \cite{kim2013}. If no natural regularization is available, it can be enforced with quadtrees \cite{chua2017}, grids \cite{grobe2020} or merged areas \cite{mcnabb2019}. Finally, Phoenixmap \cite{zhao2021} uses concave hulls for each category and encodes density along the outline. However, point densities are less of a concern for SBSS than point values.

\subsection{Parameter Space Interactions}
\label{sec:related-work--parameter}

As we present interactive visualizations to set spatial parameters, we see interactive visualizations for other parameter spaces as related work and discuss them here.
When the parameter space is multidimensional with a manageable amount of dimensions, parallel coordinate plots (PCPs) are highly popular \cite{johansson2016}. They show dimensions as parallel axes and data points in the multidimensional space are encoded as polylines. Each vertex coincides with an axis where the respective value of the dimension is. Common interactions with PCPs are reordering and brushing. PCPs have also been explored as an ideation tool to quickly create new design options \cite{mohiuddin2020}. If data points exist in multiple sets, nested PCPs \cite{wang2017} are an option. PCPs were, in an immersive environment, also combined with scatterplots into parallel planes \cite{brunhart-lupo2016}. Each plane is a scatterplot of two variables, and polylines pass through these planes. This may help when the number of dimensions grows, but they may at some point be too many. In this case, users might still insist on sliders \cite{hazarika2020} or one could persuade them to work with a dimensionally-reduced view \cite{orban2019}.

PCPs are great for multidimensional non-spatial data and have, therefore, been applied in combination with spatial data visualizations to enable multidimensional spatial data analysis \cite{gahegan2002, demsar2008, meseery2018, opach2018}. But different approaches are needed when the parameters have a spatial or temporal dimension. World Lines \cite{waser2010} is an interaction paradigm to steer a flooding simulation while it happens. At different points in time, analysts may want to, e.g., place sand bags to protect an area from water, and explore the parallel universes (with and without sand bags). It preserves this branching temporal structure in the interface. In the spatial case it is popular to provide the analyst options, e.g., in the form of a spreadsheet metaphor \cite{jankun-kelly2000}, where possible parameter settings and their effect on the outcome are arranged next to each other. In such cases, the analyst often interacts with the parameter space through a selection in the output space, like in DreamLens \cite{matejka2018}. A constrained editing mode that can optimize an objective interactively, e.g., the flight distance of a model airplane \cite{umetani2014}, is another useful interaction idiom. Finally, obtaining outputs by randomly precomputing large numbers of parameter settings \cite{sedlmair2014} may be the simplest approach, but gets less useful and more computationally expensive the larger the parameter space is.



\section{Background}
\label{sec:background}

\subsection{Data Definition}
\label{sec:data-definition}

As touched upon in \Fref{sec:related-work--spatial-statistics}, a multivariate spatial dataset in our case consists of $p$-dimensional vectors $\bm{x} \in \mathbb{R}^p$ at spatial locations $S \subseteq \mathbb{R}^2$. The vector at the $i$-th location is denoted as $\bm x (\bm s_i)$.

A parameter setting $(r, k)$ consists of a partition, or regionalization, $r$ and a point neighbourhood, or kernel, $k$. A kernel $k$ is a set of non-overlapping rings with inner and outer radius ($0 \leq c_{in}<c_{out}$). The location of a kernel does not need to be set, as a kernel will be evaluated for every location in each region. A regionalization $r$ partitions the spatial domain into a set of regions such that each location $\bm{s} \in S$ is contained in exactly one region. Hence, there are neither overlaps between regions nor leftover locations.

A kernel $k$, applied to $n$ locations $\bm s_i \in S$, defines a symmetric $n \times n$ neighbourhood matrix $\mathbf{K}$. If for the distance $d_{ij}$ between two locations $\bm s_i$ and $\bm s_j$ and any ring in $k$ $c_{in} \leq d_{ij} \leq c_{out}$ holds, the $i$-th and $j$-th row/column of $\mathbf{K}$ contain the value $1$, $0$ otherwise.


\subsection{Considerations for Selecting Parameters}
\label{sec:parameter-considerations}

There are several requirements and considerations to take into account when selecting a parameter setting $(r,k)$ for SBSS.

\paragraph*{Technical Requirements.} From assumptions in SBSS theory follows, as already touched upon in \Fref{sec:data-definition}, that the regions in $r$ must not overlap. To further simplify finding regions, we require that their union must contain all locations in $S$. These are easy to enforce automatically, but two other considerations require the human analyst.

\paragraph*{Balance Region and Kernel Size.} A guideline by our collaborators to reduce the estimation error in the weighted average (\Fref{sec:related-work--spatial-statistics}) is that each region in $r$ should contain a reasonably large amount of locations. The same is true for a kernel $k$, which should capture reasonably many locations in each region. Hence, $r$ and $k$ are not chosen independently. If a region contains sparsely distributed locations, the kernel needs to be bigger than for a denser region to capture the same number of locations. It clearly is also not useful if, e.g., the inner radius of the kernel is bigger than a region, as no locations would be captured. In practice, analysts should first select regions and kernels based on the guidelines below and afterwards verify that no region/kernel is \enquote{too small,} based on a threshold that is appropriate for the dataset and application. In our evaluations (\Fref{sec:eval-sbss-experts}), participants initially set this threshold to 5\% of data points. If too small regions/kernels are identified, analysts may proceed regardless or merge/expand regions/kernels, again following guidelines below.

\paragraph*{Reconcile With Domain Knowledge.} Another recommendation from SBSS theory is that regions should be selected such that they enclose areas where variables behave, or can be expected to behave, very differently from the other regions. This, however, depends on the concrete dataset SBSS is applied to, and prior knowledge about the physical reality it represents. As an example, if the measured variables are about air quality, it may make sense to distinguish between urban and rural regions in the data, but in case the measured variables are elements in soil, different soil types could guide the regionalization. Similarly, a kernel should be selected such that it encapsulates the spatial dependence of different latent processes in the dataset, i.e., a kernel should cover the distance within which a process may be noticeable. Such a latent process might be, e.g., emissions from driving cars, which influence air quality up to a distance of a few hundred meters \cite{liu2019f}. In the same way as a regionalization, the kernel parameter also clearly requires the domain knowledge of the analyst.
Such considerations are difficult to quantify, but may be supported by others that are easier to (data-driven considerations). For instance, which latent processes can be expected in the dataset depends on which variables were measured and how far apart. The spatial dependence of a variable, important for kernel selection, can be expressed with a variogram \cite{cressie1993m}. It is possible to automatically partition a spatial domain \cite{guo2008}, which could be an initial suggestion for this complex parameter.

To summarize, SBSS parameter selection is characterized by a small set of rules that can be easily verified automatically, and a larger fuzzier set of guidelines that require human reasoning and domain knowledge. How our visual analytics prototype supports both is the topic of \Fref{sec:the-thing}.


\section{Task Description}
\label{sec:task-abstraction}

We describe users and their tasks using the design triangle by Miksch and Aigner \cite{miksch2014}. The data is described in  \Fref{sec:data-definition}.

\paragraph*{Users.} As SBSS is a relatively novel statistical method, our users are for now SBSS experts who want to investigate their method on real data instead of the usual simulation studies. While SBSS experts have formal education in mathematics/statistics and are knowledgeable in spatial statistics, we paid attention that this is not a requirement for our visual designs. We anticipate that as interest in SBSS grows in the future, domain experts without such qualifications will require our interactive visualizations, too. Our users' main tool is RStudio, an integrated development environment (IDE) for R \cite{rcoreteam2020}, a language for statistical computing. RStudio is text-based and allocates one part of its interface to show a non-interactive visualization (which has to be programmed by the user with, e.g., ggplot2 \cite{wickham2016}).

\paragraph*{Tasks.} User tasks emerge from parameter setting considerations described earlier (\Fref{sec:parameter-considerations}). First and foremost, users need to be able to \emph{quickly and efficiently enter parameter settings} (T1), also complex ones. As can easily be imagined, this is not possible with a text interface. For this reason, users currently favor parameter settings that can be easily described with code, such as regionalizations that are grids or regular slices in a particular direction, although these may not correspond to the spatial reality in the data. Furthermore, they have to \emph{balance region and kernel size} (T2) and \emph{reconcile possible regions and kernels with their domain knowledge} (T3). The former is currently difficult as regions and kernels are selected without a direct manipulation paradigm, and the latter because only a single visualization is visible at a time.

We obtained the necessary tasks in a user-centered design process with experts in SBSS and geostatistics. We also asked an expert in geochemistry for feedback on our visualizations during the design phase. He underlined the importance of task T1, that the system should be highly interactive and allow to produce many parameter settings in little time.


\section{Visualizations and Interactions}
\label{sec:the-thing}

\begin{figure*}
    \centering
    \begin{tikzpicture}
	\node[anchor=south west,inner sep=0] (image) at (0,0)  {\includegraphics[width=\textwidth]{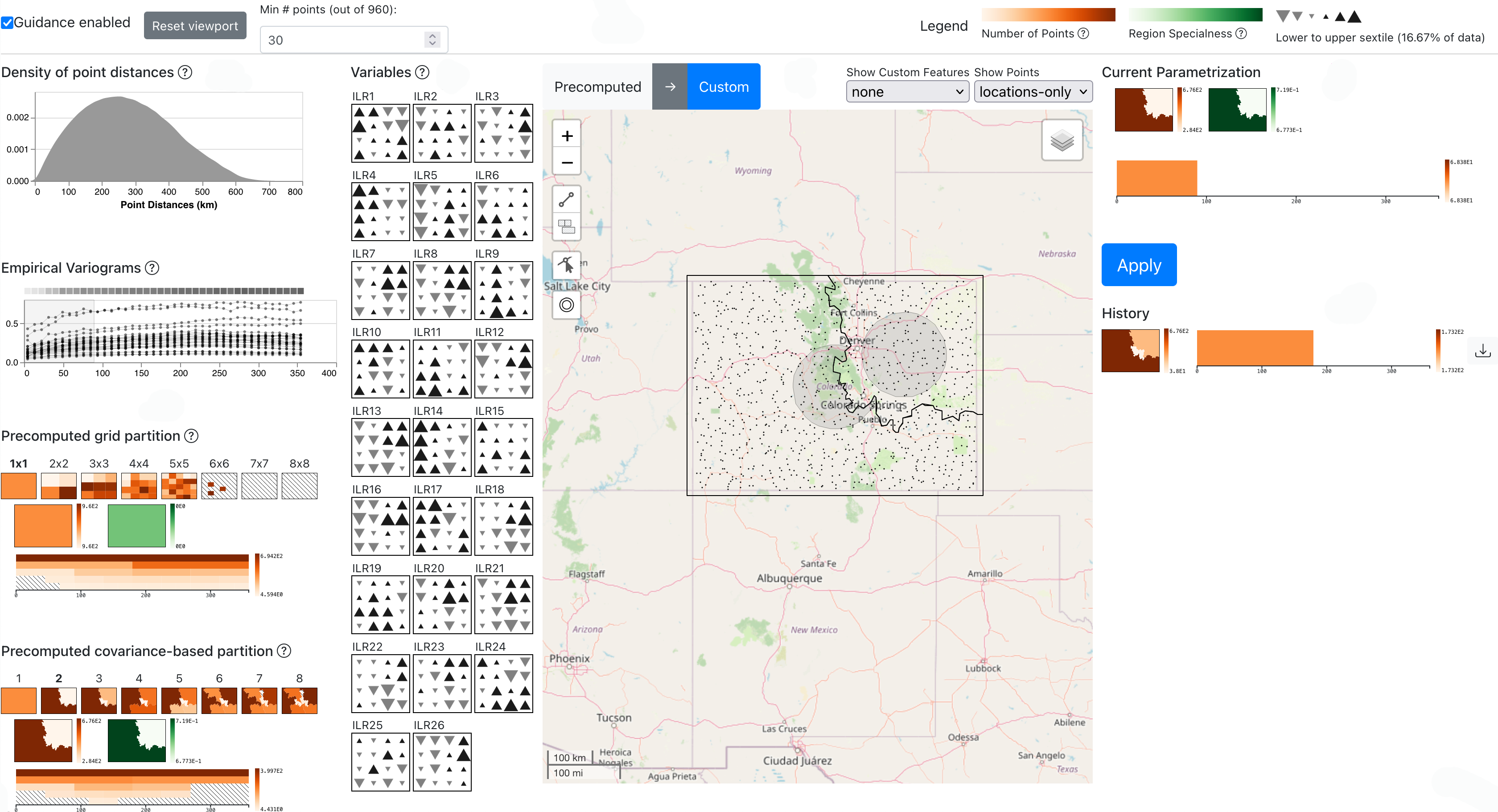}};
    \begin{scope}[x={(image.north west)},y={(image.south east)}]
		\node [textWcolor, anchor=north west] (a) at (1, 0.3) {A} ;
		\node [textWcolor] (b) at ($(a)+(110mm, 0mm)$) {B} ;
		\node [textWcolor, anchor=north west] (c) at (0.9, 0.2) {C} ;
		\node [textWcolor] (d) at ($(c)+(0mm, -20mm)$) {D} ;
		\node [textWcolor] (e1) at ($(c)+(0mm, -40mm)$) {E} ;
		\node [textWcolor] (e2) at ($(c)+(0mm, -65mm)$) {E} ;
		\node [textWcolor] (f) at ($(a)+(0mm, -7mm)$) {F} ;
		\node [textWcolor, anchor=north west] (g) at (0.9, 0.525) {G} ;
		\node [textWcolor, anchor=north west] (h) at (0.9, 0.9) {H} ;
		\node [textWcolor] (i) at ($(h)+(0mm, -25mm)$) {I} ;
    \end{scope}
    \end{tikzpicture}
    \caption{Screenshot of our prototype with the \emph{Colorado} dataset. It shows the toolbar including legends (B) and a filter for number of locations (A), visualizations supporting data-driven considerations (C and D, \Fref{sec:distance-distribution-and-variograms}), precomputed regionalizations and kernels (E, \Fref{sec:region-kernel-search}), variables as small multiples (F, \Fref{sec:spatial-summary}), an interactive map (G, \Fref{sec:interactive-map}), the analyst's current selection (H) and past selections (I).}
    \label{fig:the-thing}
\end{figure*}

In this section we describe the interactive visualizations of our prototype (\Fref{fig:the-thing}) and relate them to the task description (\Fref{sec:task-abstraction}). 
We implemented those as part of a client-server architecture with the client being a JavaScript application and the server written in R. The latter is mainly to use the \texttt{SpatialBSS} R package \cite{sbsspackage} that provides necessary functions. Both client and server carry out time-intensive computations once and re-use results, thereby allowing fluid interactions. The software is available online \cite{zotero-6809}. 
We describe and show the design with changes made \emph{after} our evaluations. We resized some elements and removed the guidance previously encoded in the blue colorscale (cf. \Fref{sec:region-kernel-quantification}).

\subsection{Interactive Map (T1, T2, T3)}
\label{sec:interactive-map}


The SBSS parameters---regions and kernels---are complex spatial objects. It is time-intensive, error-prone, and frustrating to define these in an indirect manner by textual commands. A direct manipulation interface for both of them seemed therefore promising for task T1. We achieve this in an interactive map, which supports the usual pan and zoom interactions. Not only can the analyst define regions and kernels directly in their spatial context (tasks T2, T3), with an interactive map it is also possible for us to show supporting data to guide the parameter selection (task T3).

Notably the map has two modes. One is the \enquote{precomputed} mode, which allows to view precomputed guidance suggestions (\Fref{sec:guidance}). If the analyst wants to build their own parameter setting or modify a precomputed one, they need to switch to the \enquote{custom} mode. In this mode they can split a region in two, merge two adjacent regions, and define a kernel directly in the map (\Fref{sec:interactive-map--direct-manipulation}). Any precomputed setting can be copied to the \enquote{custom} mode for modification.

\subsubsection{Visualization of Regions and Kernels (T2)}

As per common convention in cartography, we show regions as polygons. They are not filled to not occlude the underlying map tiles, which provide important information.

We show the current kernel configuration as shaded concentric circles at the geometric center of each region. This is for two reasons: First, a single kernel configuration is used for all regions, which was an acceptable simplification to our collaborators. Hence we may copy it as soon as a new region is defined. Second, there is no single center for a kernel, as it will be evaluated at all locations. Shown at a region's center we can expect that analysts will be able to reason well about a kernel and region's relative size (task T2). We use the continuity Gestalt principle to encode which region a kernel belongs to and crop the rings by the region's boundary.

\subsubsection{Direct Manipulation of Regions and Kernels (T1)}
\label{sec:interactive-map--direct-manipulation}

At first, the interactive map just shows the bounding box of all locations, and neither locations nor variables, to not clutter it from the start. This is important because we do not know in advance how many locations the dataset contains.

In discussions with our collaborators we learned that they expect regions to be coarse and few. This is partly because a region must not be too small (task T2, \Fref{sec:parameter-considerations}). As we further required all locations to be assigned to a region, a \emph{regionalization} is shaped by splitting an existing region in two along a user-defined border, which is provided by drawing a polyline through a region on the map (\Fref{fig:parameters-ixn-region}). To merge adjacent regions it is sufficient to select them. This allows to quickly define also complex regions, while maintaining correctness (task T1).

A kernel configuration is a set of concentric rings. They are defined as follows. First, the analyst picks a center point anywhere on the map. From there, the analyst has to choose alternatingly the outer and inner radius of a ring (\Fref{fig:parameters-ixn-kernel}). The process is terminated and kernel definition complete when the kernel selection mode is turned off. With this direct manipulation approach and supporting views it is easily noticeable when there are overlapping kernels. Hence, we support kernel definitions in a quick and correct way (task T1). Please refer to the video in the supplemental material for all visual feedback we provide.

\subsubsection{Additional Data (T3)}

Several additional spatial objects may be shown on the map to support selection of the parameters (task T3).

\paragraph*{Custom Annotations.} Our approach offers the ability for custom map annotations. The analyst may provide and overlay any GeoJSON feature collection \cite{butler2016}. This way, their domain knowledge can be externalized and visually encoded to support parameter selection. In a geochemical setting, it could be a soil atlas \cite{hrnciarova2009}. The cursor then snaps to the boundary of features, further simplifying the process.

\paragraph*{Locations and Variables.} Analysts may choose to show just the locations in the map encoded as points. This is a compromise between no locations and showing a spatial variable. They may, however, also overlay any single spatial variable of the dataset instead. These are encoded in the same way as in the small multiples explained in \Fref{sec:spatial-summary}: A colored triangle of differing size shows the sextile ($1/6$ or 16.67\% of the data).

\paragraph*{Base Layer.} Finally, we provide several base layers of the map to choose from. The default is OpenStreetMap and OpenTopo, Thunderforest Landscape \cite{thunderforest2021} and Satellite are also available. We expect that these cover most commonly needed information as they provide layers optimized for both rural/natural and urban areas.

\subsection{Guidance (T1, T2, T3)}
\label{sec:guidance}

The set of possible regionalizations and kernel definitions is vast and it is difficult for analysts that lack deep domain knowledge, to find a starting point. They do not know what possible parameter settings look like and how they compare. Hence, they need guidance \cite{ceneda2017}. We provide \emph{orienting} and \emph{directing} guidance in the following way (\Fref{fig:the-thing}, E). Possible regionalizations are precomputed, using a current strategy of analysts (grid-based) and one that matches SBSS experts' recommendations (covariance-based). Similarly, possible kernel settings are precomputed. We show these as suggestions (\emph{directing}, \Fref{sec:region-kernel-search}) and color-code them by quantification measures (\emph{orienting}, \Fref{sec:region-kernel-quantification}).

\subsubsection{Finding Regions and Kernels}
\label{sec:region-kernel-search}

A regionalization is visualized as choropleth map, with univariate color scales as defined in \Fref{sec:region-kernel-quantification}. We use two strategies to provide suggestions for regionalizations.

\paragraph*{Grid-based Regionalization.} For lack of better tooling, grids are currently a popular setting for the regionalization parameter. These can be quickly precomputed in a straightforward manner. We use square $n \times n$ grids with $n$ from 1 to a user-defined granularity.

\paragraph*{Covariance-based Regionalization.} Recall that in \Fref{sec:parameter-considerations} we described that regions should be selected such that the variable interactions are different. When we consider the covariance of variables as a measure, we can compute suggestions for a regionalization automatically. We first convert the point dataset to a polygon dataset using a Voronoi diagram. Then we group adjacent similar Voronoi cells using the REDCAP regionalization algorithm \cite{guo2008}. In REDCAP's terms, we use $dist_{edge}$ as edge length and $hg_r$ as region heterogenity:
\begin{align*}
dist_{edge}(\bm s_i, \bm s_j) &= \| \bm x(\bm s_i)\bm x(\bm s_i)^T - \bm x(\bm s_j)\bm x(\bm s_j)^T \|_F \\
hg_r &= \sum \| (\bm x (\bm s_i) - \overline{\bm{x}}_r)(\bm x (\bm s_i) - \overline{\bm{x}}_r)^T - \mathbf{Cov}_r \|_F
\end{align*}
$i,j$ are indices of locations, $\mathbf{Cov}_r$ is the sample covariance matrix of all locations in the region and $\overline{\bm{x}}_r$ the means of variables in the region. $\|\cdot\|_F$ denotes the Frobenius norm. With these hyperparameters for REDCAP, we gain regionalizations for a user-defined number of regions. This approach was very successful in our evaluation (\Fref{sec:evaluation}).

\begin{figure}
    \centering
    \includegraphics[width=0.66\linewidth]{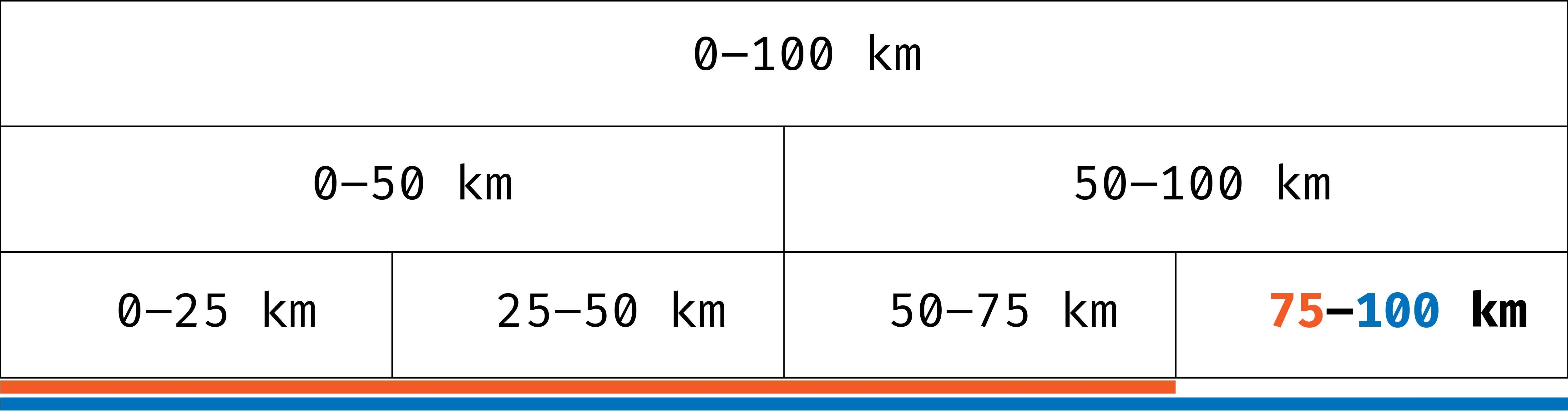}
    \caption{Structure of the visualization for kernels (\Fref{fig:the-thing}, E). Each kernel is a rectangle representing a part of some largest kernel.}
    \label{fig:kernel}
\end{figure}

\paragraph*{Kernels.} For kernel suggestions we consider only kernels with a single ring, as the rings have no influence on each other. We obtain smaller rings by a recursive binary partition of a largest ring. To visualize the precomputed rings, we show them as stacked bars (\Fref{fig:kernel}), where the Y axis encodes ring thickness and the X axis distance. The left edge of a bar marks the inner radius of a ring, the right edge the outer radius. The bars are colored according to a color scale described in \Fref{sec:region-kernel-quantification}. Single rings can be selected to be viewed on the map, but any combination of rings may be defined manually.

\subsubsection{Quantifying Regions and Kernels}
\label{sec:region-kernel-quantification}

\paragraph*{Number of Locations.} For analysts it is important to know how many locations are contained in a region and captured by a kernel (\Fref{sec:parameter-considerations}). Hence, for any region in a regionalization, we simply count the number of locations in it. For a kernel in a region, we compute the neighbourhood matrix $\mathbf{K}$ (\Fref{sec:data-definition}) and define the number of locations captured by the kernel as the mean of column sums in $\mathbf{K}$. As one column in $\mathbf{K}$ contains the neighbourhood for a single location defined by a kernel, it is the average neighbourhood size. This metric is encoded in the orange color scale. It may be used to inform, e.g., if a region may be split further or if two adjacent regions should be merged (tasks T2, T3).

\paragraph*{Insufficient Number of Locations.} A pattern of diagonal stripes appears when the number of locations in a region or kernel neighbourhood is smaller than a custom threshold. This way, analysts can easily detect too small regions or kernels (task T2).

\paragraph*{Region Covariance Difference.} In \Fref{sec:parameter-considerations} we outlined that regions should be selected such that the variable interactions are different. One way to describe those are by the sample covariance matrix $\mathbf{Cov}_r$ of all $\bm x (\bm s_i)$ in a region. The difference of each $\mathbf{Cov}_r$ to the global sample covariance matrix $\mathbf{Cov}$ can then the quantified by the Frobenius norm: $\|\mathbf{Cov}-\mathbf{Cov}_r\|_F$. Higher values indicate more locally different variable interactions. This metric is encoded in the green color scale. This should be used to identify as many as much locally different regions as possible, as long as they are also reasonable for a domain expert (task T3).

\paragraph*{Eigenvalue Difference.}

SBSS theory states that high quality recovering of the latent field is achieved if the eigenvalues of the local covariance matrices (\Fref{sec:related-work--geospatial-vis}) evaluated on the \emph{latent} field are as different as possible \cite{bachoc2020,muehlmann2021}. Hence, a promising parameter setting maximizes the difference between these eigenvalues.
Unfortunately, the latent field is unknown beforehand.
However, in this spirit, our collaborators suggested that the eigenvalue difference of the local covariance matrices evaluated on the \emph{input data} might be a useful metric to suggest the latent field recovery quality of a parameter setting. In the version of the prototype we used for our evaluations (\Fref{sec:evaluation}), this metric was encoded in a blue color scale. As it was not well accepted among study participants, probably due to its unreliability, we removed it from the final design presented in this paper (\Fref{fig:the-thing}).

\subsection{Summary of One Spatial Variable (T1)}
\label{sec:spatial-summary}

We decided to show all involved variables separately to the interactive map as small multiples (\Fref{fig:the-thing}, F). Following parameter selection considerations in \Fref{sec:parameter-considerations}, analysts need to identify areas of the spatial domain in which many variables have consistent values, which makes it necessary to show all variables at once. This is effectively a manual regionalization (\Fref{sec:region-kernel-search}) and the small multiples simultaneously provide an overview of all variables.

A single spatial variable is summarized by aggregating it to a grid. The size of the grid can be interactively changed by the analyst (semantic zoom). For each grid cell, the median value of the variable is encoded by a triangle symbol showing the percentile it falls in (\Fref{fig:the-thing}, B). This design was preferred by our collaborators over a heatmap or isocontours. We divide the data in sextiles ($1/6$ or 16.67\% of the data). The lower three sextiles are gray and upside-down triangles, the upper three are black and upright triangles. This double encoding is redundant, but allows to perceive contiguous regions due to the shared color and also intuitively indicates which percentiles are shown: Downward-pointing triangles show lower values, upward-pointing triangles higher values. As for our collaborators the extreme values are of interest, values away from the 3rd and 4th sextiles are shown with bigger triangles. The relative sizes were chosen based on \cite{dent1996}.


\subsection{Distance Distribution and Variograms (T3)}
\label{sec:distance-distribution-and-variograms}

We show two plots to support the data-driven selection considerations (\Fref{sec:parameter-considerations}).

\paragraph*{Distance Distribution.} To know how far away locations are from each other we show a density plot of all pairwise distances (\Fref{fig:the-thing}, C). From that an analyst can easily see if the spatial scale of the dataset is on hundreds of meters or thousands of kilometers. This is in addition to the interactive map (\Fref{sec:interactive-map}).

\paragraph*{Variograms.} The empirical variogram is an established plot in spatial statistics \cite[Chapter~2]{cressie1993m} that shows the spatial dependence of a variable, i.e., how its value changes with increased distance. With the (binned) distance on the X axis, the Y axis encodes the average squared difference between any point pair whose distance falls in that particular bin. We combine variograms of all variables in the dataset by superpositioning them (\Fref{fig:the-thing}, D). In \Fref{sec:parameter-considerations} we explained that kernels can be selected such that they encapsulate dissimilar spatial behavior of variables. To support this assessment, we add a grayscale-coded square on top of each bin that encodes the variance. Hence, darker squares point to bins with more dissimilar spatial behavior. When the analyst selects a kernel, its current extent is interactively shown in the variogram view.



\section{Evaluation}
\label{sec:evaluation}

In previous sections we described users and their tasks (\Fref{sec:task-abstraction}) and presented our interactive visualizations (\Fref{sec:the-thing}). In this section we describe our efforts to evaluate these visualizations. We were interested in the following research questions:

\begin{itemize}
    \item (RQ1) Do our interactive visualizations enable more \emph{efficient} parameter selection? I.e., can analysts enter complex settings in less time?
    \item (RQ2) Do our interactive visualizations enable more \emph{effective} parameter selection? I.e., do they allow analysts to enter previously-impractical settings?
    \item (RQ3) Is our designed guidance effective, i.e., is it semantically meaningful and accepted by users?
\end{itemize}

Evaluations were carried out with three groups of participants. We presented our prototype to five visualization experts (\Fref{sec:eval-vis-experts}), who judged its value using a questionnaire \cite{wall2019}. This is to confirm that we did not make gross mistakes in the visualization design phase. After that, we invited two external SBSS experts (\Fref{sec:eval-sbss-experts}), who did not take part in the design phase, to a user study. Here we were interested in how our prototype can improve their parameter selection process. Finally, we showed the covariance-based regionalization guidance (\Fref{sec:guidance}) and latent dimensions (output from a parameter setting made by the second author) to an expert in geochemistry (\Fref{sec:eval-domain-experts}), to judge how meaningful suggested partitions and acquired results are.

Hence, we combined quantitative and qualitative approaches. However, we did not deem it useful to compare RStudio and our visualizations in a quantitative way involving time and error. The two are based on completely different interaction paradigms and provide wildly differing levels of support to the analyst. From the discussion with SE2 we think we were right in that decision.

\paragraph*{Datasets Used in Evaluation.} For visualization experts we exclusively used the \emph{GEMAS} \cite{reimann2014} dataset (2\,108 locations / 18 variables), because it covers most of Europe and we expected it therefore to be somewhat relatable. SBSS experts preferred the \emph{Kola moss} \cite{reimann1998} (594 / 31) and \emph{Colorado} \cite{smith2010} dataset (960 / 27). For guidance judgement we again used the \emph{GEMAS} and \emph{Kola moss} datasets, because the domain expert is one author of them and intimately familiar. All datasets are publicly available.

\subsection{SBSS Experts}
\label{sec:eval-sbss-experts}

Regarding our research questions of efficiency and effectivity, we interviewed two people who work a lot in RStudio and are SBSS experts. We introduced them (SE1 and SE2), who have at least one publication on SBSS, to our visualizations. They used the prototype on a dataset they chose. These were different datasets. We asked them to produce a few parameter settings using our prototype. We did not provide any requirements to this task, to not constrain their exploration and ideas. Of course, we helped them if they did not remember visual encodings or interactions. We asked them to vocalize their plans and intentions (\enquote{think aloud}). After they were done or the time ran out, we discussed the visualizations and interactions in an unstructured fashion. The sessions took around 75~minutes each. SE2 even provided us beforehand parameter settings they made in RStudio.

In the beginning, SE1 had difficulties using the prototype. Especially the distinction between the \enquote{precomputed} view-only map mode and the \enquote{custom} editable mode was confusing, as both looked similar but edit controls were missing in one of them. However, after 15–20~minutes, SE1's interactions became quite fluid. SE2 had no problems from the start. This suggests that there is little training time necessary to use our prototype. 

Both experts, being knowledgeable about SBSS but not the application domain of geochemistry, relied in their selection process heavily on the guidance of our prototype.

\paragraph*{SE1} browsed through many suggestions, but had trouble to commit to any particular setting. It is possible that we provided too many options, or at least should not have shown them all at once. For lack of better judgement, SE1 settled for four parameter settings from our guidance system, with minor modifications.

\paragraph*{SE2} went about it in a more structured way, but produced only a single setting in the end. At first SE2 also mostly browsed through (grid-based) suggestions and inspected them in the interactive map. SE2 also paid attention to the colormaps of the suggestions, although more on the orange and green one. At some point SE2 decided to find the most locally different region and clicked through grid-based regionalizations. These tended to show regions in the center as darkest, to confirm this SE2 inspected the variable summaries. There SE2 noticed that many variables had consistently similar values in the top right square and in the left-most column of the map. We pointed out that this observation matches the covariance-based regionalizations, and SE2 used this guidance more from that point on. SE2 combined the covariance-based maps and the variable summaries to decide for one regionalization, specifically the most fine-grained one that did not split the regions of interest identified earlier. Then a process of fine-tuning began, where SE2 split and merged regions to distribute locations evenly, while keeping as much of the baseline regionalization intact as possible. Finally, kernels were selected with support of the variograms. In the end, SE2 selected a parameter setting with much more complex regions than in prior attempts made using RStudio. Judging from our conversations with the domain expert (\Fref{sec:eval-domain-experts}), this setting is likely more realistic, too.

While we had plenty of time with SE2, it was not the case with SE1, with whom we were not able to discuss drawbacks and benefits deeply. SE1 raised no improvement suggestions for the prototype, but mentioned that the purpose of the Eigenvalue guidance in the blue colormap is not clear. This was reinforced by SE2, who also did not look at it that much. We believe that this is because it relates only to the output and its impact on that is unclear. Therefore, this guidance is unreliable and we removed it in the final prototype. Regarding using our prototype vs. RStudio, SE1 mentioned that \enquote{the precomputations are extremely useful.} This was also echoed by SE2, but suggested that it would have been nice if similar suggestions existed for the kernel parameter, too. SE2 also noted that being able to see the original variables would have been useful. This is related to a technical detail with geochemical data: Since these are measured as part of a whole (e.g., mg/kg in a soil sample), it is necessary to apply some data transformations first \cite{aitchison1982} and our prototype showed variables only after these transformations. We suggested SE2 to recreate one of their existing parameter settings with our prototype. However, SE2 declined with an interesting answer: It would \enquote{probably be faster} but pointless, as they would \enquote{very likely not be interested in choosing the same settings,} given the fewer constraints and additional supporting views of our prototype. We take this as strong evidence for our initial assumption, that providing tailored interactive visualizations change the selection process, and for our first two research questions. As both experts made use of our guidance suggestions, we see this as supporting evidence also for our third research question, that our guidance is effective.


\subsection{Visualization Experts}
\label{sec:eval-vis-experts}

We asked visualization experts to judge our visualization design. While good design does not automatically entail a more efficient/effective selection process, bad design most likely prohibits it. We used the heuristic value of visualization approach by Wall et al. \cite{wall2019} (ICE-T), because it is a good compromise between insight gained for us and time required for participants. We introduced five visualization experts from two universities, who are Ph.D. students or postdocs in visualization, to the SBSS problem domain and our prototype (\Fref{sec:the-thing}). Five experts are sufficient according to the power analysis by Wall et al. The experts were allowed to use the prototype on their own and ask as many questions as necessary, until they felt confident enough to fill out the questionnaire. We discussed the terminology beforehand. The sessions took around one hour each and were conducted solely by the first author. The results are depicted in \Fref{tab:ice-t}.

\begin{table}[]
\centering
\begin{tabular}{@{}lrr@{}}
\toprule
\multicolumn{1}{c}{\textbf{Component}} & \multicolumn{1}{c}{\textbf{Mean}} & \multicolumn{1}{c}{\textbf{Std.dev.}} \\ \midrule
Insight                                & 6.35                                 & 0.98                                     \\
Time                                   & 6.52                                 & 0.65                                     \\
Essence                                & 6.05                                 & 0.89                                     \\
Confidence                             & 6.00                                 & 2.98                                     \\ \bottomrule
\end{tabular}
\caption{Results of the ICE-T evaluation with visualization experts.}
\label{tab:ice-t}
\end{table}

It can be seen that our approach was rated very well across ICE-T components. For our purpose, we see Time and Insight, in that order, as the most important components, which also were rated highest on average. Wall et al. \cite{wall2019} state that a visualization design can be considered successful if the mean score is greater than five, which we clearly achieved. We provide the raw questionnaire results as supplementary material.

However, the standard deviation in the Confidence component is very high. The reason for this is that two out of four statements were often either deemed not applicable or rated badly by our participants. These pertain to facilitating learning more broadly about the data domain and helping to understand data quality. The former is not important to our prototype, as it is designed to support parameter selection only, and not general data analysis capabilities. The latter partly is: SBSS requires complete data, therefore good data quality can be seen as a precondition. On the other hand, duplicate or outlier entries could still exist, but would be invisible due to occlusion and the percentile summary. This could be solved, e.g., by a layout that avoids occlusion, by annotations when occlusion happens or by highlighting symbols if their value is greater than a user-defined number of standard deviations.


Several points were raised in the open discussion. The variogram was unknown to all participants, but thought to be a good way to show spatial dependence of variables. That there should be more space for the map, also because the other views could be shown conditionally, was raised by two participants. Our triangle symbols were deemed both intuitive and not ideal because it is not easy to see where the center is. Participants also suggested to make it possible to analyze custom groups of variables.

\subsection{Domain Expert}
\label{sec:eval-domain-experts}

\begin{figure}
     \centering
     \begin{subfigure}[b]{0.2775\textwidth}
         \centering
         \includegraphics[width=\textwidth]{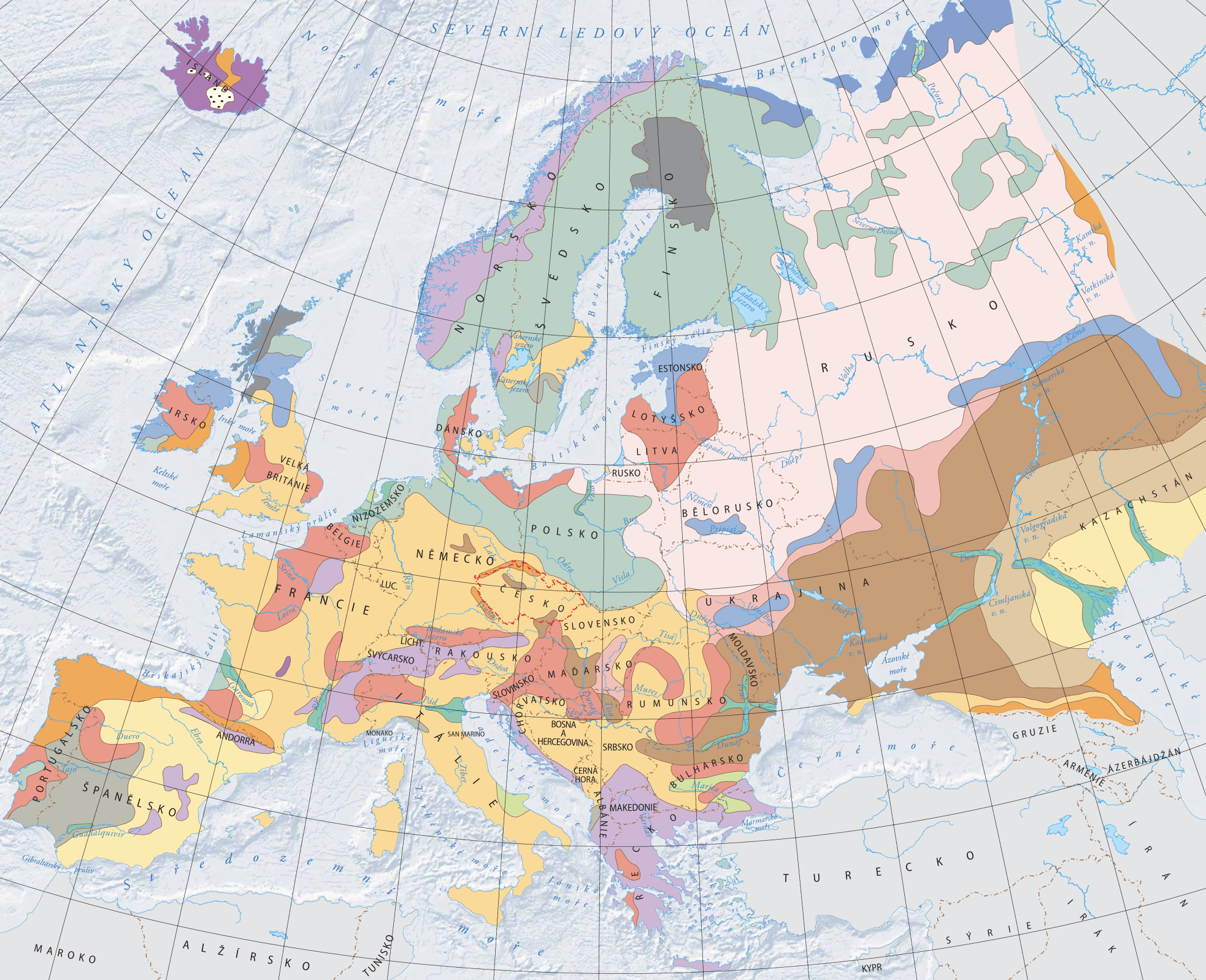}
         \caption{Main soil types in Europe. Reproduced from \cite{hrnciarova2009}.}
         \label{fig:guidance-vs-atlas-atlas}
     \end{subfigure}\hfill
     \begin{subfigure}[b]{0.1625\textwidth}
         \centering
         \includegraphics[width=\textwidth]{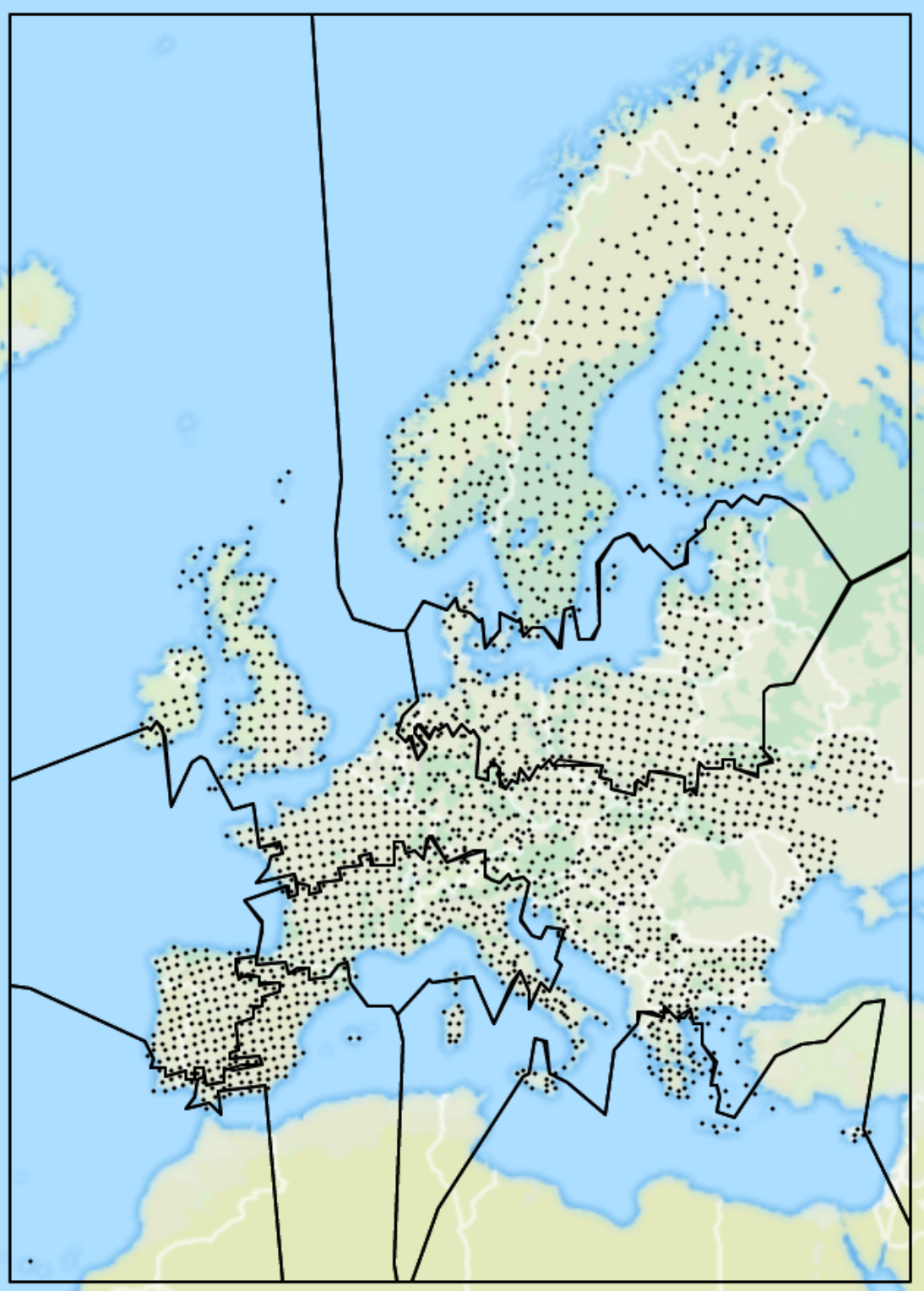}
         \caption{8 automatic regions for the \emph{GEMAS} dataset.}
         \label{fig:guidance-vs-atlas-guidance}
     \end{subfigure}
    \caption{Comparison of a) a map of soil types in Europe and b) our regionalization guidance.  While not perfect, as it is limited here to 8 regions and the dataset captures many more latent processes than just soil type, our guidance suggests similar boundaries, such as East/West Spain or North/South Baltic region.}
    \label{fig:guidance-vs-atlas}
\end{figure}

We showed the automatic covariance-based regionalizations for the \emph{GEMAS} dataset (\Fref{fig:guidance-vs-atlas-guidance}) to the geochemistry expert. He immediately recognized geologically and, therefore, geochemically distinct areas that are characterized by their soil (\Fref{fig:guidance-vs-atlas-atlas}), such as eastern Spain (Calcisol), Central Europe (Cambisol), Southern Baltic region (Albeluvisol), or the Nordic countries (Podzol). He further mentioned that such an automatic regionalization based on multivariate data would likely be helpful for geologists and geographers as an initial estimation of homogeneous regions. This is often necessary, e.g., because non-spatial methods, like PCA, must not be used on inhomogeneous data \cite[Chapter 14]{reimann2008}. The domain expert was not able to reconcile the automatic regionalizations with known processes in the \emph{Kola moss} dataset. In our opinion, even this negative assessment is useful, as it suggests a stationary SBSS setting (i.e., no regionalization required). Overall, we take this as evidence that our regionalization suggestions can reflect real processes and be a starting point even for domain experts.

\begin{figure}
     \centering
     \begin{subfigure}[t]{0.45\linewidth}
         \centering
         \includegraphics[width=\textwidth]{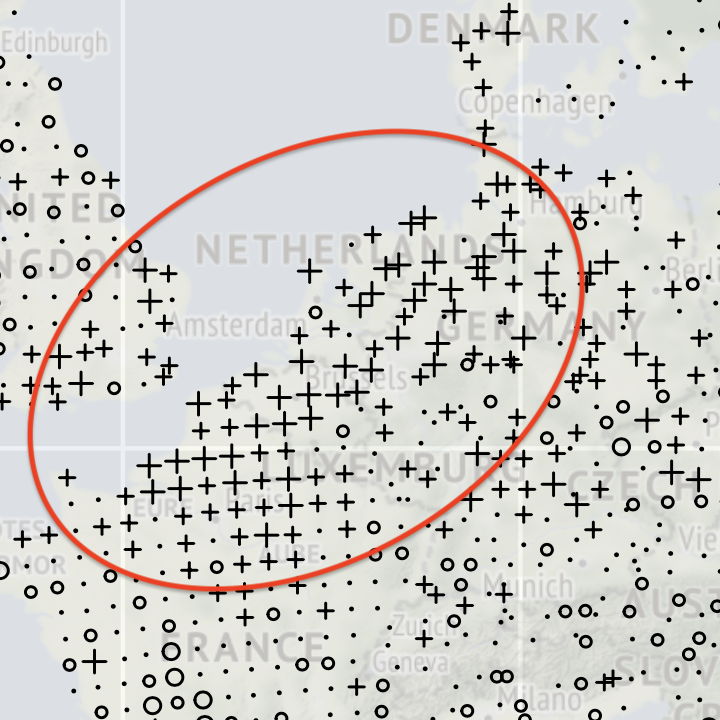}
         \caption{Unidentified process causes pattern in France and Germany.}
         \label{fig:gerfrabel}
     \end{subfigure}\hfill
     \begin{subfigure}[t]{0.45\linewidth}
         \centering
         \includegraphics[width=\textwidth]{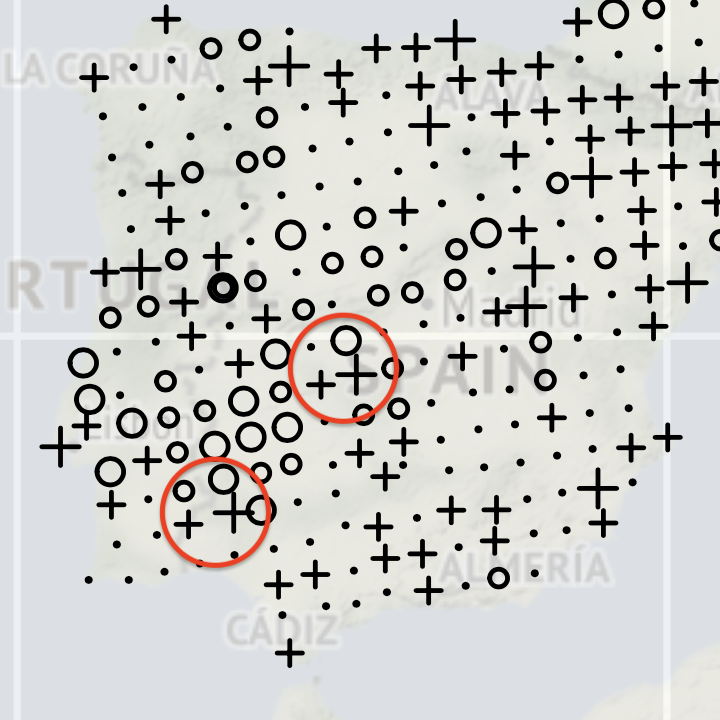}
         \caption{Patterns near mercury (top circle) and copper mines (bottom).}
         \label{fig:riotinto}
     \end{subfigure}
     \caption{Insights into the \emph{GEMAS} dataset with SBSS. Images show high (crosses) and low (circles) values of a latent dimension \cite{reimann2008}. Zoom, crop and red annotation by the authors.}
\end{figure}

To further test the applicability of our interactive visualizations, the second author used them to define a parameter setting on the \emph{GEMAS} dataset. This is difficult because of the dataset's complexity (2\,108~locations covering Europe, 18 variables), especially for someone unfamiliar with the application domain. Yet it took him only a few minutes. We then plotted static maps of the resulting latent components and showed them to the geochemistry expert, who noticed familiar, surprising and unknown patterns. Unexpected was a structure in the area of North France, Belgium and Germany (\Fref{fig:gerfrabel}). The expert speculated it is caused by sediments, but then the pattern would extend east- instead of westwards. While there are possible explanations, like population density, more research is necessary to confirm them. More unexpected patterns were identified near known sites of mining activity in Seville (Rio Tinto) and Almadén (\Fref{fig:riotinto}). This was insofar surprising to the expert as Almadén is a mercury deposit and Rio Tinto copper/zinc, yet neither mercury nor copper were part of the dataset we used. The expert generally was impressed that a lot of known processes, like historic geological events (e.g., Oslo rift, glacial period), were so well visible, even though our dataset did not include the \enquote{most interesting elements.} Revealing the same interesting patterns with fewer variables has monetary implications in geochemistry, as some elements are expensive to measure within useful detection limits (and could be excluded). These insights show how useful SBSS can be for multivariate analysis of spatial data, and how accessible they became to novice users with our interactive visualizations.


\subsection{Limitations and Discussion}

Being a research prototype, our VA approach does have its limitations (see also \Fref{sec:eval-sbss-experts} and \Fref{sec:eval-vis-experts}). The computational demand increases with the number of locations in the dataset ($\mathcal{O}(n^2)$ per regionalization with REDCAP), hence the precomputations may take several minutes. Further, any region currently must be a single contiguous area without holes.

Our research questions pertained to the efficiency (RQ1) and effectivity (RQ2) of parameter settings with our interactive visualizations and the effectiveness of the guidance we designed (RQ3).
To answer RQ1 and RQ2 we performed, on the one hand, a heuristic evaluation with visualization experts. Our prototype scored particularly well in the Time component, as participants strongly agreed that it provides efficient interactions. Visualizations were also deemed appropriate, except to find data quality issues. The latter is a minor issue as, in practice, SBSS expects a properly preprocessed dataset.
On the other hand, we introduced our prototype to two external SBSS experts, who used it to select parameters on a dataset of their choice. Little training time was necessary and the visualizations and guidance suggestions were considered useful. One expert stressed how our prototype allows to set previously practically impossible parameter settings. Therefore, we think RQ1 and RQ2 can be answered positively.
Our third research question (RQ3) was about the effectiveness of our guidance. The availability of regionalization suggestions was considered very useful by SBSS experts. A novice in geochemistry (the second author) was quickly able to select parameter settings that lead to surprising insights for a domain expert. We therefore think that also this research question can be answered positively.

\section{Conclusion}
\label{sec:conclusion}

 SBSS is a desirable tool for multivariate spatial data analysis. It requires setting complex spatial tuning parameters: a partition of the spatial domain (regionalization) and a spatial neighbourhood configuration (kernels). In this paper, we presented a visual-interactive prototype that supports and guides analysts in finding appropate settings, thereby rendering it more usable in practice. We developed it in close collaboration with experts in SBSS, geostatistics, and geochemistry. The prototype contains several interactive capabilities to modify parameters and guiding visualizations. We evaluated the prototype quantitatively using a heuristic evaluation with five visualization experts and qualitatively with two external SBSS experts, who were not part of the design process, and a geochemistry expert. Our evaluations show that

\begin{itemize}
    \item our visualizations are appropriate and the prototype allows highly interactive exploration of possible parameter settings,
    \item our prototype allows SBSS and visualization experts to select parameters more flexibly, efficiently, and realistically,
    \item our guidance suggestions can be semantically meaningful to a domain expert and are considered helpful by SBSS experts.
\end{itemize}




During this study we discovered partial results that we think can be transferred between the domains of geostatistics, geochemstriy, and visual analytics for mutual benefit. For instance, the ideas of variograms and regionalizations rarely occur in visual analytics literature for spatial data. Flexible and interactive variograms \cite{haslett1991} and regionalizations are useful for exploratory analysis of spatial data, which in turn can support geostatistical modeling. It would be interesting to further explore, how these can be combined with state-of-the-art interactive visualizations. Similarly, automatic regionalizations may be useful for geochemists, as they suggest homogeneous areas from multivariate data, which are interesting by themselves and can be analyzed by other methods, such as PCA. To further improve upon the concept of regionalizations it would be beneficial to make them uncertainty-aware, as we learned that region boundaries in practice may not be crisp and clear-cut when multiple influencing processes overlap.

Our contributions present a first step towards the effective practical use of SBSS. In the future we could look into more quantitative approaches at several stages of the design study, e.g., comparing SBSS results obtained with our prototype to a ground truth, investigating an objective-oriented parameter selection approach, or conduct experiments to find out which visualizations are best for the tasks we identified.
Further topics arise as a result from our focus on visual-interactive parameter \emph{selection}. Because several parameter settings are tried in practice, it raises the question how common visual parameter analysis tasks, like sensitivity analysis, can be possible with spatial parameters. Finally, it would be beneficial if the exploration of multiple SBSS results are supported by interactive visualizations.

\section*{Acknowledgements}

This work was funded by the Austrian Science Fund (FWF) under grant P31881-N32. We sincerely thank Clemens Reimann for his support and advice. We also thank Helwig Hauser and our anonymous study participants for their time and valuable discussions.


\printbibliography

\end{document}